\documentclass[12pt]{article}
\pdfoutput=1
\usepackage[utf8x]{inputenc}
\usepackage{hyperref}
\usepackage{graphicx}
\usepackage{subcaption}
\graphicspath{{./picture/}}
\usepackage{empheq}
\usepackage{amssymb}
\usepackage{mathtools}
\usepackage{commath}
\usepackage{verbatim}
\usepackage[dvipsnames]{xcolor}
\usepackage{soul}
\usepackage{dsfont}
\usepackage{booktabs}
\usepackage{amsmath}
\usepackage{hhline}
\usepackage[scale={0.75,0.8}]{geometry}
\usepackage{url}
\usepackage{caption}
\usepackage[numbers, sort&compress]{natbib}
\usepackage{epsfig}
\usepackage{lscape}
\usepackage{mathrsfs}
\usepackage{nicefrac} 
\usepackage{upgreek}
\usepackage{bbm}
\usepackage{psfrag}
\usepackage{hyperref}
\usepackage{hyperref}

\newcommand{\iu}{{\rm i}}
\newcommand{\Mu}{\mu}
\newcommand{\M}{\bar{M}}
\newcommand\ii{{\rm i}}		
\usepackage{slashed}				
\usepackage{feynmp-auto}

\newcommand{\Uai}{U_{ai}^2}
\newcommand{\Ua}{U_{a}^2}
\newcommand{\Ui}{U_{i}^2}
\newcommand{\U}{U^2}

\newcommand{\muup}{{\mu^\prime}}

\begin{document}

\title{ 
\vspace{-2cm}
\begin{flushright}
{\scriptsize CP3-19-35}
\end{flushright}
\vspace{0.5cm}
{\bf \boldmath On Lepton Number Violation in Heavy Neutrino Decays at Colliders} \\[8mm]}
\date{}

\author{Marco Drewes,$^{a}$ Juraj Klarić,$^{b}$ Philipp Klose$^{a}$ \\ \\
{\normalsize \it$^a$Centre for Cosmology, Particle Physics and Phenomenology,}\\{\normalsize \it Universit\'e catholique de Louvain, Louvain-la-Neuve B-1348, Belgium}\\ 
{\normalsize \it$^b$ 
Institute of Physics, Laboratory for Particle Physics and Cosmology (LPPC),} \\ {\normalsize \it 
Ecole Polytechnique F\'{e}d\'{e}rale de Lausanne (EPFL), CH-1015 Lausanne, Switzerland
}\\
} 

\maketitle
\thispagestyle{empty} 
\begin{abstract}
  \noindent 
We study the perspective to observe lepton number violating signatures from heavy Majorana neutrino decays at colliders in view of the requirement to explain the light neutrino masses via the seesaw mechanism. 
In the minimal model with only two heavy neutrinos and in the $\nu$MSM one can identify three distinct regions in the mass-mixing plane. 
For Majorana masses above the electroweak scale the branching ratio for lepton number violating processes at the LHC is generically suppressed.
For masses well below the electroweak scale that are probed in displaced vertex searches or at fixed target experiments lepton number violation is the rule and can only be avoided at the cost of fine tuning. 
In between there is a mass regime where both possibilities coexist.
In models with more than two heavy neutrinos the larger parameter space allows for more freedom, but our results remain qualitatively correct unless there is a mass degeneracy amongst more than two of the heavy neutrinos.
\end{abstract}

\tableofcontents

\section{Introduction}

Right handed neutrinos $\nu_{R}$ appear in many popular extensions of the Standard Model (SM) of particle physics. 
As SM gauge singlets, the $\nu_{R}$ can have a Majorana mass term $\overline{\nu_{R }} M_M \nu_{R}^c$ that breaks lepton number $L$ in the SM. 
Depending on the values of their Majorana mass(es) they may explain various phenomena in particle physics and cosmology, cf. e.g. \cite{Drewes:2013gca} for a review.
Most importantly, the mixing between the "sterile" neutrinos $\nu_{R}$ and the SU(2) charged left handed SM neutrinos $\nu_{L}$ can generate small masses for the standard model neutrinos 
via the \emph{type I seesaw mechanism}~\cite{Minkowski:1977sc, GellMann:1980vs, Mohapatra:1979ia, Yanagida:1980xy, Schechter:1980gr, Schechter:1981cv}.
In addition the $\nu_{R}$ could explain the baryon asymmetry of the universe via \emph{leptogenesis} during their decay \cite{Fukugita:1986hr} or production \cite{Akhmedov:1998qx,Asaka:2005pn}
for a wide range of values of their Majorana masses (including experimentally accessible ones \cite{Chun:2017spz}), 
provide a viable Dark Matter candidate \cite{Dodelson:1993je,Shi:1998km}, or explain the anomalies observed in some neutrino oscillation experiments \cite{Abazajian:2012ys}.

The mixing between a sterile neutrino flavour $i$ and the SM generation $a = e, \mu, \tau$ can be quantified by  small angles $\theta_{ai} \ll 1$.
In terms of the matrix of heavy neutrino Yukawa couplings $F$ to the SM lepton doublets, the Higgs field expectation value $v$, and the Majorana mass $M_M$, the mixing angles $\theta_{ai}$ and the light neutrino Majorana mass matrix $m_\nu$ at tree level read\footnote{
The mass and mixing matrices are often expressed in terms of the Dirac matrix $m_D= v F$ as  $m_\nu^{\rm tree} = - m_D M_M^{-1} m_D^T$ and $\theta=m_D M_M^{-1}$.
}
 \begin{eqnarray}\label{eq:seesaw}
 \theta_{ai}=v (F M_M^{-1})_{ai}
\quad ,
 \quad
 m_\nu^{\rm tree} = - v^2 F M_M^{-1} F^T
= - \theta M_M \theta^T.
\end{eqnarray}
While the sterile neutrinos $\nu_R$ are gauge singlets, the mass eigenstates after electroweak symmetry breaking $N_i$ participate in all weak processes with amplitudes that are suppressed by $\theta_{ai}$,%
\footnote{The physical mass eigenstates $N_i$ are given by the quantum mechanical admixture $N_i \simeq  \nu_{R i} +  \theta_{ai}\nu_{L a}^{c} + \text{c.c.}$ between the $\nu_R$ and $\nu_L$. 
Here $\text{c.c.}$ denotes the charge conjugation that e.g.~acts as $\nu_L^c=C\overline{\nu_L}^T$ with $C=\ii\gamma_2\gamma_0$ in the Dirac and Weyl representations.}
which makes it possible to search for them experimentally \cite{Shrock:1980ct,Shrock:1981wq}.%
\footnote{We refer the reader to the reviews in refs.~\cite{Atre:2009rg,Deppisch:2015qwa,Lindner:2016bgg,Cai:2017mow,Antusch:2016ejd,Beacham:2019nyx,Alimena:2019zri} for details on the perspectives to find heavy neutrinos experimentally.}

This has triggered an increasing interest in searches for $N_i$ with masses below the TeV scale at colliders and fixed target experiments.
The Majorana mass term $M_M$ can violate the SM lepton number $L$ and lead to lepton number violating (LNV) decays of the heavy neutrinos.
These have long been considered to be a golden channel for their discovery in collider experiments because they have low SM backgrounds, which makes them experimentally very appealing.
The ATLAS \cite{Aaboud:2018spl}, CMS \cite{Sirunyan:2018xiv} and LHCb \cite{Aaij:2014aba} experiments have all published limits on the properties of heavy neutrinos from searches for LNV. 
However, it has been pointed out in ref.~\cite{Kersten:2007vk} that the branching ratios of LNV processes at colliders can be parametrically suppressed in the pure type-I seesaw model, which could potentially limit the discovery potential of such searches. 
The argument can be summarised as follows: 
\begin{itemize}
\item 
The angles $\theta_{ai}$ control both, the heavy neutrino contribution to the light neutrino mass matrix \eqref{eq:seesaw}
\emph{and} their production cross section at colliders $\sigma_{N_i} \sim |\theta_{ai}|^2\sigma_{\nu_a}$, where $\sigma_{\nu_a}$ is the light neutrino production cross section. 
The seesaw relation \eqref{eq:seesaw} naively predicts $|\theta_{ai}|^2 \sim m_i/M_i < 10^{-9} M_i/{\rm GeV}$, meaning that the heavy neutrinos are either too heavy or too feebly coupled to be produced at colliders.
Much larger mixing angles than this can only be achieved if there are cancellations in the matrix valued equation \eqref{eq:seesaw}, which for a generic choice of parameters would require fine-tuning~\cite{Gluza:2002vs}.
\item 
Comparably large $|\theta_{ai}|^2$ can be made consistent with the observed small neutrino masses $m_i$ without tuning
 if the $\nu_{R i}$ approximately conserve a generalised lepton number $\bar{L}$ \cite{Shaposhnikov:2006nn,Kersten:2007vk}. 
Approximate $\bar{L}$ conservation is in fact the only way to achieve this that is technically natural, i.e. stable under radiative corrections \cite{Moffat:2017feq}.
\item 
The conservation of $\bar{L}$ would automatically suppress the S-matrix elements for all LNV processes, along with other LNV observables such as neutrinoless double $\beta$ decay.
\end{itemize}
As a result one may conclude that LNV processes can never be observed at colliders in technically natural implementations of the seesaw mechanisms.

A number of ways have been pointed out to avoid this conclusion.
Most straightforwardly, the suppression of $\sigma_N$ can be avoided if the heavy neutrinos have other, new interactions in addition to the $\theta$-suppressed weak force. 
We do not discuss this possibility here; reviews that cover this topic can be found e.g. in refs.~\cite{Deppisch:2015qwa,Cai:2017mow}, cf.~also ref.~\cite{Dev:2013oxa} for an explicit example.
Instead we focus on the question whether the small $\bar{L}$ violation that must necessarily occur to generate non-zero Majorana masses for the light neutrinos is sufficient to give LNV signatures in collider searches for heavy neutrinos 
that exclusively interact through their mixing with the ordinary neutrinos in technically natural scenarios.  
In recent years different signatures of LNV at colliders have been studied \cite{Cvetic:2010rw,Cvetic:2015naa,Cvetic:2015ura,Dib:2016wge,Anamiati:2016uxp,Das:2017hmg,Dib:2017iva,Antusch:2017ebe,Antusch:2017pkq,Cvetic:2018elt,Hernandez:2018cgc,Cvetic:2019rms,Abada:2019bac,Gluza:2015goa,Gluza:2016qqv,Pilaftsis:1997dr,Bray:2007ru}.
In the present work we want to address the question what the perspectives are to see any of these signatures in view of constraints from neutrino oscillation data.
To give a quantitative answer to this question we consider two criteria,
\begin{itemize}
\item[1)] 
There is a lower bound on the amount of $\bar{L}$ violation from the requirement to explain the experimental fact that the light neutrino masses $m_i$ are not exactly vanishing.
\item[2)] 
There is an upper bound on the amount of $\bar{L}$ violation from the requirement to explain the smallness of the $m_i$ without fine tuning. 
That is, the breaking of the underlying symmetry must be small enough for the $m_i$ to be stable under radiative corrections.
\end{itemize}
Applying these criteria allows one to divide the parameter plane spanned by the heavy neutrinos mass and mixing into a region where the branching ratio of LNV processes is of order unity,
another region where it is parametrically suppressed, and an intermediate region in which both possibilities coexist.

\section{Lepton number violation in experiments: general considerations}

\paragraph{Neutrino masses and approximate lepton number conservation.}
For the purpose of collider searches, the most important properties of the heavy neutrinos are their masses $M_i$ and the mixing angles $\theta_{ai}$ which determine the suppression of their weak interactions.
To quantify the overall suppression of the heavy neutrino interactions we introduce the notation
\begin{align}
	\Uai \equiv |\theta_{ai}|^2\,, && \Ui  \equiv \sum_a |\theta_{ai}|^2 \,,\\\notag
	\Ua \equiv \sum_i |\theta_{ai}|^2\,, && \U  \equiv \sum_{a,i} |\theta_{ai}|^2\,.
\end{align}
The \emph{seesaw relation}  \eqref{eq:seesaw} connects these quantities to the light neutrino masses, which are given by the square roots of the eigenvalues of $m_\nu^\dagger m_\nu$.
Large mixing angles $\theta_{ai}$ can be consistent with the light neutrino masses if there are cancellations among the different entries in this matrix.
Such cancellations may appear tuned if they are accidental, but
they occur in a systematic and technically natural manner if the Yukawa matrix $F$ and Majorana mass $M_M$ respect 
a generalised lepton number $\bar{L} = L + L_{\nu_R}$ \cite{Shaposhnikov:2006nn,Kersten:2007vk}, where $L_{\nu_R}$ is a charge that can be assigned to the sterile neutrinos.
Approximately $\bar{L}$ conserving models can predict mixing angles $\theta_{ai}$ that are large enough to give observable $N_i$ production cross sections in experiments while respecting the relation \eqref{eq:seesaw} without fine tuning. 
For instance, for $M_i$ below the electroweak scale and $|F_{ai}|$ of the order of SM charged lepton Yukawa couplings, the $N_i$ are easily within reach of searches at the LHC \cite{Sirunyan:2018mtv,Aad:2019kiz}. 
In this regime they can also explain the observed baryon asymmetry of the universe \cite{Abada:2018oly}. 
Models that can accommodate a symmetry of this kind e.g. include \emph{inverse seesaw} type scenarios~\cite{Wyler:1982dd,Mohapatra:1986aw,Mohapatra:1986bd,Bernabeu:1987gr} and the \emph{linear seesaw}~\cite{Akhmedov:1995ip,Akhmedov:1995vm} as well as scale invariant models~\cite{Khoze:2013oga}
and the \emph{Neutrino Minimal Standard Model} ($\nu$MSM)~\cite{Asaka:2005an,Asaka:2005pn}.

If the $\bar{L}$ conservation is exact, then the heavy neutrinos must either decouple (have vanishing Yukawa couplings) or be organised in pairs $N_i$ and $N_{j}$ with $j=i+1$ and \cite{Moffat:2017feq}
\begin{eqnarray}\label{eq:symmetry}
F_{a j} = \ii F_{a i} 
\quad , \quad 
M_i = M_j 
\end{eqnarray}
leading to $\theta_{a j} = \ii\theta_{a i}\equiv \theta_a$. 
In this limit the light neutrino masses $m_i$ vanish identically and the heavy neutrinos that couple to the SM are of the Dirac type.\footnote{In fact, in the symmetric limit there are no Majorana fermions in the Lagrangian that couple to the SM: 
the $\nu_{L a}$ are massless Weyl fermions, and each pair $\nu_{R i}$ and $\nu_{R j}$ merge into a single Dirac fermion $\psi_N=\frac{1}{\sqrt{2}}(\nu_{R i} + \nu_{R i}^c) + \frac{\ii}{\sqrt{2}} (\nu_{R j} + \nu_{R j}^c)$. We use four component chiral spinors with $\nu_{R,L}=P_{R,L} \nu_{R,L}$ to represent the neutrinos.}
The $\bar{L}$ conservation conveniently suppresses the rate of neutrinoless double $\beta$ decay, which otherwise would impose a very strong upper bound on $|\theta_{ei}|$ \cite{Bezrukov:2005mx,Blennow:2010th}. 
At the same time the underlying symmetry implies that the eigenvalues of $M_M$ come in degenerate pairs, which is favourable for resonant leptogenesis \cite{Pilaftsis:2003gt} or leptogenesis in the $\nu$MSM \cite{Asaka:2005pn,Canetti:2012kh}.
The downside is that it also leads to a parametric suppression of the matrix elements for
all other LNV processes.  
At first sight this suggests that it is hopeless to observe LNV processes at colliders:
The only way to make the $\theta_{ai}$ large enough to obtain sizeable production cross sections at colliders while keeping the neutrino masses small is to impose a protecting symmetry, but the very same symmetry parametrically suppresses the branching ratios of LNV processes
relative to the lepton number conserving (LNC) processes.
However, the symmetry cannot be exact because otherwise the light neutrinos would be exactly massless. 
The relevant question is therefore:
What is the relation between the small $\bar{L}$ violation that gives rise to non-zero neutrino masses and the branching ratio for LNV heavy neutrino decays at colliders?

\paragraph{LNV from heavy neutrino oscillations.}
An example of  a LNV signal is the detection of two leptons with the same sign. The rate for such an event is computed from the Feynman diagram\footnote{
In the present work we assume that the collision energy is sufficient to produce the heavy neutrinos as on-shell particles. In this case the diagram \eqref{eq:ss2lep} dominates and the behaviour inside the detector can be understood analogously to the well-established light neutrino flavour oscillations, but with much shorter oscillation and decay lengths due to the smaller boost factor and (much) shorter lifetimes.  For $M_i$ above a few hundred GeV the $N_i$ cannot be produced on-shell at the LHC, and the LNV is mediated via exchange of virtual particles. In that case the simple treatment below does not apply, but the perspectives for an experimental discovery are in any case much worse.
}
\begin{fmffile}{samesignDilepton}
\setlength{\unitlength}{0.22mm}
\fmfset{arrow_len}{3mm}
\fmfset{zigzag_width}{1thick}
\fmfcmd{%
style_def
majorana expr p =cdraw p;cfill (harrow (reverse p, .5));cfill (harrow (p, .5))
enddef;
style_def
alt_majorana expr p =cdraw p;cfill (tarrow (reverse p, .55));cfill (tarrow (p, .55))
enddef;
}
\begin{align}\label{eq:ss2lep}
	\Gamma_{X\rightarrow \ell_a \ell_b} \sim
\left\vert
\sum_i
\quad
\parbox{31.68mm}{
\begin{fmfgraph*}(144,89)
	\fmfstraight
	\fmfleftn{i}{6}
	\fmfrightn{o}{6}
	\fmf{plain,tension=6}{i3,v1,i4}
	\fmf{zigzag,tension=6}{v1,v2}
	\fmf{fermion}{v2,v5}
	\fmf{phantom}{v5,o6}
	\fmf{majorana,tension=1.5,label=$N_i$,label.side=right}{v2,v3}
	\fmf{fermion}{v3,o4}
	\fmf{zigzag,tension=3}{v3,v4}
	\fmf{plain,tension=3}{o1,v4,o2}
	\fmfv{label=$\theta_{ai}$,label.angle=90}{v2}
	\fmfv{label=$\theta_{bi}$,label.angle=90}{v3}
\end{fmfgraph*}
}
\quad
\right\vert^2\,.
\end{align}
\end{fmffile}
In the limit \eqref{eq:symmetry} the contributions from the two members $N_i$ and $N_{j}$ of the $i$-th pair
to this amplitude exactly cancel, which is simply a manifestation of the fact that lepton number is conserved in this limit.%
\footnote{Strictly speaking it is $\bar{L}$ that is conserved, not the SM lepton number $L$. 
However, searches for $L$ violation in fully reconstructed final states are the most reliable probe of $\bar{L}$ violation.
Whenever the $N_i$ decay back into SM particles in the detector, the SM lepton number is conserved in the final state if $\bar{L}$ is conserved. 
If they leave the detector, then $L$ is violated in the observed particles in the detector. 
To conclude from this that $L$ is violated in nature would require to distinguish this process from one where $L$ has been carried away by a SM neutrino, which is practically very difficult. 
} 
However, the cancellation crucially relies on the coherence of the quantum state during the process. 
It has been pointed out by various authors that the heavy neutrinos undergo oscillations in the detector \cite{Boyanovsky:2014una,Cvetic:2015ura,Anamiati:2016uxp,Dib:2016wge,Das:2017hmg, Antusch:2017ebe,Antusch:2017pkq, Cvetic:2018elt,Hernandez:2018cgc,Cvetic:2019rms} if there are deviations from the relation \eqref{eq:symmetry}. 
These can effectively destroy the coherence of the quantum state because they practically randomise the phase if the heavy neutrinos undergo many oscillations during their lifetime. 
The deviations must be parametrically small for the $\bar{L}$  symmetry to explain the smallness of the neutrino masses without fine tuning, 
meaning that a straightforward computation of the matrix element from the diagram \eqref{eq:ss2lep} yields a non-zero, but parametrically small result.\footnote{
Intuitively one may expect that the branching ratio for LNV decays is also helicity suppressed if the particles are relativistic. 
However, while such this argument in good approximation holds for light neutrinos \cite{Kayser:1982br},
it turns out that this only leads to a different angular distribution when it comes to the decay of heavy neutrinos, 
cf.~e.g.~ref.~\cite{Arbelaez:2017zqq, Balantekin:2018ukw}.
One way to understand this is that the loss of definite chirality (governed by the heavy neutrino masses $M_i$) always happens quicker than the flavour oscillations (governed by the physical splitting $\Delta M_{\rm phys}$ between their masses).
} 
This parametric suppression can, however, be overcome by a separation of scales between the frequency of the heavy neutrino oscillations and their lifetime. A quantitative treatment requires the use of resummed propagators and solving the equations of motion in real time rather than computing the S-matrix perturbatively.

For the minimal model with $n=2$ that we study in section \ref{Sec:benchmark}
small perturbations around \eqref{eq:symmetry} that can be characterised by 
\begin{align}\label{SymmetryBreaking}
	F =
	\begin{pmatrix}
		F_{e} (1+ \epsilon_e)    & \ii F_{e} (1-\epsilon_e)\\
		F_{\mu} (1+ \epsilon_\mu)  & \ii F_{\mu} (1-\epsilon_\mu)\\
		F_{\tau} (1+ \epsilon_\tau) & \ii F_{\tau} (1-\epsilon_\tau)
	\end{pmatrix}
	\ , \
	M_M =
	\begin{pmatrix}
		\M (1+\Mu) & 0\\
		0 & \M (1-\Mu)
	\end{pmatrix}
\end{align}
with $\Mu, \epsilon_a \ll1$. 
This explicit form can also be used to study interference within one pseudo-Dirac pair in scenarios with $n>2$. 
Interference between members of different pairs  is usually not relevant because those resonances in general lie on different mass shells. We discuss the important exception that there is a mass degeneracy between different pairs at the end of section \ref{Sec:benchmark}.


The frequency of heavy neutrino oscillations in the laboratory frame is roughly given by $\omega\sim(M_j^2-M_i^2)/(2 E_N)$, where $E_N$ is the heavy neutrino energy.
The coherence of the quantum state is effectively destroyed if the heavy neutrinos undergo many oscillations during their lifetime $\uptau_N$, i.e., $\tau_N \omega \gg 1$.
The lifetime is $\uptau_N \sim \gamma/\Gamma_N$ with the Lorentz factor $\gamma=E_N/M_i$ and the heavy neutrino decay width $\Gamma_N$.%
\footnote{In principle the decay widths of the individual heavy neutrinos are different. 
Moreover, the mass and interaction basis of the heavy neutrinos are in general not identical because $M_M$ and $F^\dagger F$ can in general not be diagonalised simultaneously. 
When the lifetime and oscillation frequency are of the same order one has to solve a matrix valued equation of motion that involves correlations between the different species. 
However, due to the approximate symmetry \eqref{eq:symmetry} the different damping rates in that equation are all of the same order, and we can use a single parameter $\Gamma_N$. 
}
For relativistic neutrinos with similar masses the condition $\tau_N \omega \gg 1$  translates into $\Gamma_N / \Delta M_{\rm phys}\ll 1$. Here $\Delta M_{\rm phys} = M_j - M_i$ the physical mass splitting between the two heavy neutrinos. 
A more precise criterion has e.g. been defined in ref.~\cite{Deppisch:2015qwa,Anamiati:2016uxp}, where it was found that the branching ratio of the opposite and same sign dilepton events is then given by the quantity\footnote{This derivation assumes that many oscillations happen within the detector volume. Highly boosted heavy neutrinos could in principle change this conclusion. However, the boost factors expected at future colliders should not exceed $\mathcal{O}(100)$, which cannot compete with the short oscillation length of the heavy neutrinos which scales as $\sim 1/{\Delta M_\mathrm{phys}}$ and is typically smaller than $\mathcal{O}(10^{-4})\mathrm{m}$.}
\begin{equation}\label{RllDef}
\mathcal{R}_{\ell \ell} = \frac{\Delta M_{\rm phys}^2}{2 \Gamma_N^2 + \Delta M_{\rm phys}^2}.
\end{equation}
The simple criterion $M_{\rm phys}=\Gamma_N$ is reproduced for 
$\mathcal{R}_{\ell \ell} = 1/3$, which we will use as a criterion to distinguish suppressed from unsuppressed LNV branching ratios in the following.

For $M_i$ well below the electroweak scale one can estimate $\Gamma_N \sim 11.9 U_i^2 \M^5 G_F^2/(96\pi^3)$ \cite{Gorbunov:2007ak}.
This implies that LNV signals are suppressed if
\begin{eqnarray}\label{subEWcond}
\frac{\Delta M_{\rm phys}}{\M} < 4 \times10^{-3} \M^4 G_F^2 U_i^2
\quad
{\rm or} 
\quad
\frac{\Delta M_{\rm phys}}{{\rm eV}} < 5\times10^{-4}\frac{\M^5}{{\rm GeV}^5} U_i^2.
\end{eqnarray}
Experimental constraints on $\theta$ in this mass range \cite{Drewes:2015iva} imply that $\Delta M_{\rm phys}$ would have to be smaller than the light neutrino mass splitting.
For heavy neutrino masses above the electroweak scale one can approximate
$\Gamma_N\simeq g^2 \sum_a |\theta_a|^2 \M^3/(32\pi m_W^2) + \mathcal{O}(m_W^2/\M^2)$.
This roughly yields the condition for the suppression of LNV
\begin{eqnarray}\label{superEWcond}
	\frac{\Delta M_{\rm phys}}{\M} < 4\times 10^{-3} \frac{\M^2}{m_W^2} U_i^2
\quad
{\rm or} 
\quad 
\frac{\Delta M_{\rm phys}}{{\rm eV}} < 6\times 10^{2} \frac{\M^3}{{\rm GeV^3}}U_i^2.
\end{eqnarray}

\paragraph{
Lower limit on $\mathcal{R}_{\ell\ell}$ from the Higgs mechanism.
}%
Since there is practically no direct experimental constraint on $\Delta M_{\rm phys}$ the question whether or not one can expect to see LNV processes can only be addressed indirectly and by theoretical arguments.
By looking at the parameterisation \eqref{SymmetryBreaking} one is tempted to draw the conclusion that $\Delta M_{\rm phys} \sim \mu\M$. 
While all theories that realise the approximate $\bar{L}$ symmetry have in common that $\Mu\ll1$, the precise range of values that one would expect for this parameter is very model dependent. 
This would mean that one cannot make any generic statement about the observability of LNV processes. In particular, the value $\Mu=0$ is perfectly consistent with light neutrino oscillation data as long as $\epsilon_a\neq0$. 

Luckily it is not correct that the physical heavy neutrino mass splitting is directly proportional to $\Mu$. The reason is that the heavy neutrinos do not only have a Majorana mass $M_M$, 
but also receive a mass from the Higgs mechanism that is of the same order as the masses of the light neutrinos. 
Their physical masses after electroweak symmetry breaking are not given by the entries of $M_M$, but by the square roots of the eigenvalues of the matrix $M_N^\dagger M_N$ with \cite{Shaposhnikov:2008pf}%
\footnote{Here we have neglected loop corrections for simplicity. In the symmetric limit the must have the same flavour structure as the tree level contribution from the Higgs mechanism.}
\begin{equation}\label{MNdef}
M_N  = M_M + \frac{1}{2} (\theta^\dagger \theta M_M + M_M^T \theta^T\theta^*).
\end{equation}
The $\mathcal{O}[\theta^2]$ term is known to play an important role for the generation of lepton asymmetries below the electroweak scale \cite{Shaposhnikov:2008pf,Canetti:2012kh}, but is usually neglected in phenomenological studies because it is much smaller than $\M$. 
It does, however, have an important effect on $\Delta M_{\rm phys}$ and thus on the oscillations in detectors.

The physical mass spitting $\Delta M_{\rm phys}$ therefore receives contributions from two sources. On the one hand there is the difference between the eigenvalues of $M_M$ that is characterised by $\Mu$, on the other hand there is the term in \eqref{MNdef} from the Higgs mechanism.
By comparing it to the seesaw relation \eqref{eq:seesaw} it is straightforward to see that the latter is parametrically of the same order $\mathcal{O}[\theta^2\M]$ as the light neutrino masses. 
One would therefore expect that $\Delta M_{\rm phys}$ is at least of the order of the light neutrino masses unless the two contributions to the
splitting between the eigenvalues of $M_N^\dagger M_N$ conspire to cancel each other.
Such a cancellation would be tuned unless it is dictated by an additional symmetry.
The previously discussed $\bar{L}$-symmetry suppresses each of the two contributions to $\Delta M_{\rm phys}$  individually, but provides no reason why the splitting between the eigenvalues of $M_N^\dagger M_N$ should be smaller than that of $m_\nu^\dagger m_\nu$.
Moreover, the contributions depend in a non-trivial way on the generally unrelated physical quantities $M_M$ and $F$ as well as $v$. 
A cancellation in $\Delta M_{\rm phys}$ would therefore necessarily involve a symmetry that relates the Yukawa couplings and Majorana masses to the properties of the Higgs boson.
In summary, one can generically expect that $\Delta M_{\rm phys}$ is at least of the same order as the observed light neutrino mass splittings.

\paragraph{
Upper limit on $\mathcal{R}_{\ell\ell}$ from stability under radiative corrections.
}
The tree level relation \eqref{eq:seesaw} imposes no constraints on the mass splitting $\Delta M_{\rm phys}$.
Instead, a limit on the size of the mass splitting between the heavy neutrinos can be derived from the requirement that the $m_i$ are stable under radiative corrections to the relation \eqref{eq:seesaw}, 
which have e.g. been studied in refs.~\cite{Pilaftsis:1991ug,Kersten:2007vk,Roy:2010xq,Lopez-Pavon:2015cga,Moffat:2017feq, Dev:2013oxa}.
The leading order radiative correction to $m_\nu$ is given by~\cite{Pilaftsis:1991ug}
\begin{align}
	m_\nu^{1-\text{loop}} = - \frac{2}{(4\pi v)^2} \theta l(M_N^2) M_N \theta^T\,,
	\label{eq:rad}
\end{align}
where the loop function is
\begin{align}
	l(x)=\frac{x}{2}\left( \frac{3 \ln (x/m_Z^2)}{x/m_Z^2-1} + \frac{\ln (x/m_H^2)}{x/m_H^2-1}\right)\,.
	\label{loopfunc}
\end{align}
If we expand the mass matrix into a part proportional to the identity matrix and a remainder, $M_N \simeq \bar{M} + \Delta M_N$, we can approximate the correction as
\begin{align}\label{LoopExpansion}
	m_\nu^{1-\text{loop}} \approx \frac{2 {l}(\bar{M}^2)}{(4\pi v)^2} m_\nu^{\text{tree}} - \frac{\M^2}{v^2}\frac{4 {l}^\prime(\M^2)}{(4\pi)^2} \theta \mathrm{Re} \Delta M_N  \theta^T\, + \mathcal{O}(\Delta M_N^2).
\end{align}
For large mixing angles, the second term in \eqref{LoopExpansion} dominates and we can parametrically estimate the size of the correction as
\begin{align}
	\mathrm{Tr}\,(m_\nu^{1-\text{loop}} {m_\nu^{1-\text{loop}}}^\dagger)\sim
	\left(\frac{4 \bar{M}^2 {l}^\prime(\bar{M}^2)}{(4\pi v)^2} U^2 ||\Delta M_N||\right)^2 
	\sim \left(\frac{4 \M^2 {l}^\prime(\bar{M}^2)}{(4\pi v)^2} U^2 \Delta M_{\rm phys} \right)^2 .
	\label{largetheta2radiative}
\end{align}
The requirement that this correction remains smaller than the tree level contribution imposes an upper bound on $\Delta M_{\rm phys}$ for given $\M$ and $U^2$. 

Here we have used rough parametric estimates for the matrix valued equations, the precise upper and lower bounds on the LNV parameters depend on the number $n$ of right handed neutrinos that are added to the SM.
In the following we discuss these bounds for the specific case $n=2$, which is the smallest number that allows to explain the observed light neutrino oscillation data.

\section{Benchmark model with two heavy neutrinos}\label{Sec:benchmark}

It has been shown in ref.~\cite{Moffat:2017feq} that heavy right handed neutrinos with Yukawa couplings $F_a$ that are large enough to give sizeable production cross sections $\sigma_N$ at colliders must be organised in pairs that obey the relations \eqref{SymmetryBreaking} with $\Mu, \epsilon_a \ll 1$. 
In the scenario with two heavy neutrinos ($n=2$) there is only one such pair.
This is the smallest number of heavy neutrinos that is required to explain the light neutrino oscillation data. 
In addition it provides an effective description for the collider phenomenology of the $\nu$MSM because
the third heavy neutrino in that model is too feebly coupled to make a significant contribution to the generation of light neutrino masses \cite{Boyarsky:2006jm}.
In the following we use this scenario as a benchmark model and comment on the case $n>2$ further below.

\begin{figure}
	\centering
	\includegraphics{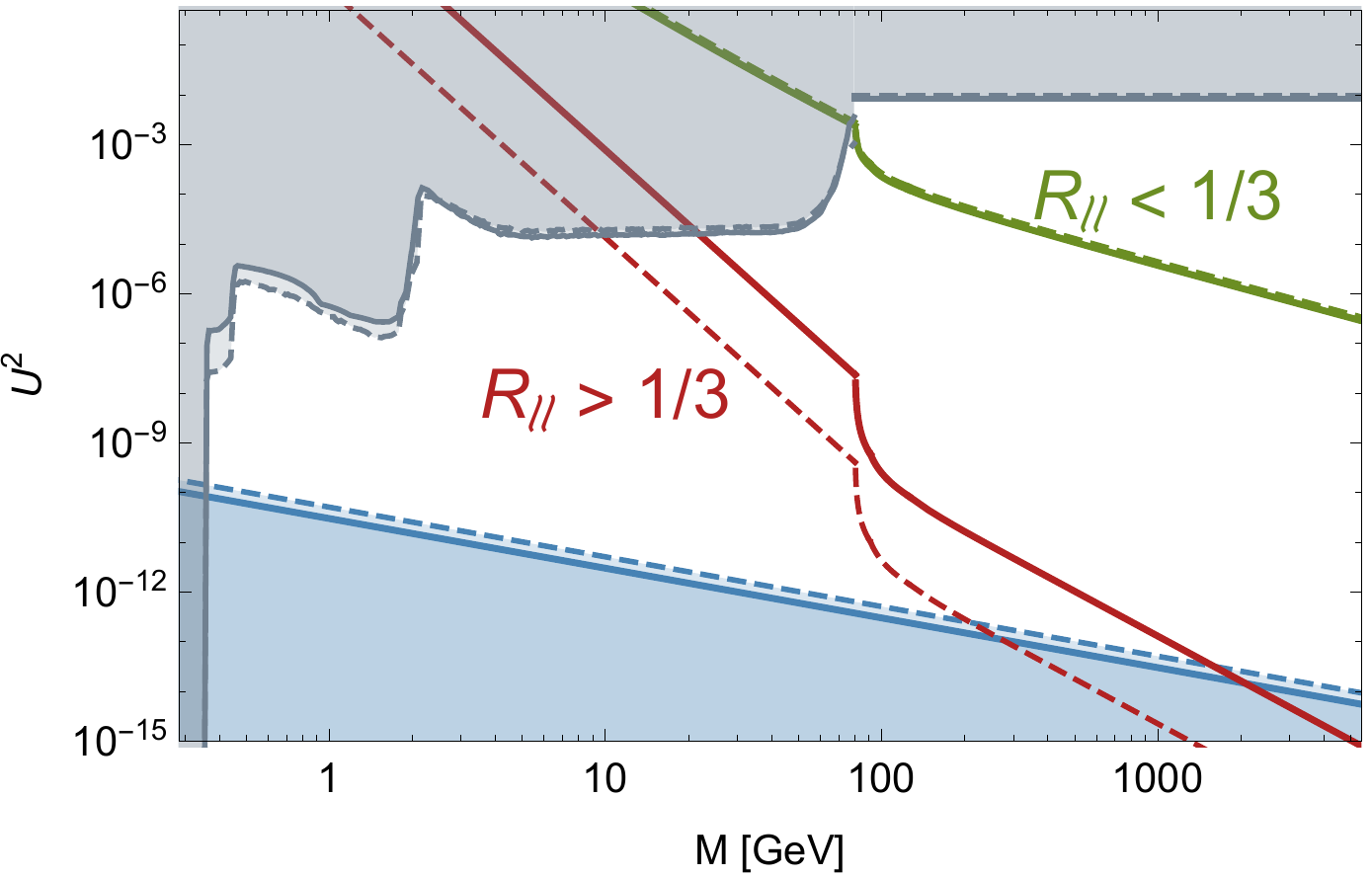}
	\caption{Parameter regions where generic parameter choices yield $\mathcal{R}_{\ell\ell}<1/3$ (above the green line)
	or $\mathcal{R}_{\ell\ell}>1/3$ (below the red line) in the minimal model with $n=2$.
	Within these regions a deviation from this generic behaviour is only possible for fine tuned parameter choices.
	The two regions are separated by a third regime where both possibilities can coexist without fine-tuning.
	The solid and dashed lines apply to normal and inverted ordering of the light neutrino masses, respectively.
Note that the suppression applies to the branching fraction of lepton number violating decays, which is proportional to the quantity $\mathcal{R}_{\ell\ell}$ defined in \eqref{RllDef}, not to the total number of LNV events, which is proportional to $\sigma_N \mathcal{R}_{\ell\ell}$.
If all final state masses are negligible and $\Gamma_N \ll \Delta M_{\rm phys}$ one can approximate $\sigma_N \propto U^2$ and  $\mathcal{R}_{\ell\ell}\propto U^{-4}$, so that the number of LNV events is quadratically suppressed by the mixing angle above the red lines. 
The shaded areas mark the regions that are excluded by experiments, based on the
global scans in refs.~\cite{Drewes:2016jae,Antusch:2014woa}.
In the gray region constraints from various direct searches at colliders and fixed target experiments dominate, cf.~e.g.~refs.~\cite{Atre:2009rg,Antusch:2014woa,Drewes:2015iva} for a discussion.
In the blue region the mixing is too small to explain the observed values of the light neutrino masses, cf.~e.g.~ref.~\cite{Drewes:2019mhg} for a discussion.
For this plot we used the simple analytic estimates for $\Gamma_N$ that are given in ref.~\cite{Atre:2009rg}, more refined computations can e.g. be found in refs.~\cite{Gorbunov:2007ak,Canetti:2012kh,Bondarenko:2018ptm,Pascoli:2018heg}.
}
	\label{fig:LNVboundary}
\end{figure}

\paragraph{Lower bound from the Higgs mechanism.}
To fix the light neutrino masses $m_i$ and parameters in the light neutrino mixing matrix $U_\nu$ to the observed values we employ the Casas-Ibarra parameterisation \cite{Casas:2001sr} for $\theta$,
\begin{equation}\label{CasasIbarra}
\theta_{ai} = \iu (U_\nu)_{aj} \sqrt{m_j/M_i} \mathcal R_{ji}.
\end{equation}
Here $\mathcal R$ is an arbitrary matrix with the property $\mathcal{R}\mathcal{R}^T =\mathds{1}$ that can be parameterised by a single complex angle $\omega$. The imaginary part of $\omega$ controls the degree of symmetry, small values of $\epsilon_a$ are obtained in the limit of large $|{\rm Im}\omega|$.
In terms of the Casas-Ibarra parameters the physical mass splitting reads \cite{Drewes:2016jae}
\begin{equation}\label{eq:DeltaMphys}
\Delta M_{\rm phys} = \sqrt{4\Mu^2\M^2  \ +  \ \Delta M_{\theta\theta}^2 \  - \  4 \ \Mu \M \  \Delta M_{\theta\theta} \ \cos(2{\rm Re}\omega)}.
\end{equation}
Here $\Delta M_{\theta\theta}=m_2-m_3$ for normal ordering and $\Delta M_{\theta\theta}=m_1-m_2$ for inverted ordering, and $2\Mu\M$ is the splitting of the eigenvalues in the Majorana mass matrix $M_M$.
In order to realise a physical heavy neutrino mass splitting $\Delta M_{\rm phys}$ that is much smaller than the splitting $\Delta M_{\theta\theta}$ between the light neutrino masses, a careful adjustment of the phase ${\rm Re}\omega$ in the Yukawa couplings and $\Mu\M$ is required.
This cancellation would require a conspiracy between physical quantities that are not related in any obvious way because it  depends on the Higgs field value $v$ through $\Delta M_{\theta\theta}$.
We can therefore confirm that $M_{\rm phys}$ is generically at least of the same order as the light neutrinos mass splittings, as argued on general grounds by comparing \eqref{MNdef} to \eqref{eq:seesaw} in the previous section.
If we set $\Delta M_{\rm phys} = \Delta M_{\theta\theta}$ in 
the ratio \eqref{RllDef} we can identify a line in the mass-mixing plane below which $\mathcal{R}_{\ell\ell} > 1/3$, which is shown in figure \ref{fig:LNVboundary}.
It can  be reproduced by setting $\Delta M_{\rm phys} = \Delta M_{\theta\theta}$
in the estimates \eqref{subEWcond} and \eqref{superEWcond}.

\paragraph{Upper bound from radiative corrections.}
The requirement that the light neutrino masses are stable under radiative corrections \eqref{largetheta2radiative}
can be used to  impose an upper bound on $\mathcal{R}_{\ell\ell}$ for given $\M$ and $\U$.
If we insert the Casas-Ibarra parametrisation \eqref{CasasIbarra} for the mixing angles $\theta$ into equation~\eqref{LoopExpansion}, we find that the radiative corrections take the form
\begin{align}
	\label{ns2radiativecorr}
	\mathrm{Tr}\,\left(m_\nu^{1-\text{loop}} {m_\nu^{1-\text{loop}}}^\dagger \right) &\approx
	\left( \mu \frac{\M^2}{v^2}\frac{4 l^\prime(\M^2)}{(4\pi)^2} \right)^2 \frac{(\sum_i m_i)^2}{2} \exp(4 \mathrm{Im} \omega) \\\notag
	&\approx \left( \mu \frac{\M^2}{v^2}\frac{4 l^\prime(\M^2)}{(4\pi)^2} \right)^2 \times 2 \bar{M}^2 U^4
\end{align}
in the limit $|\mathrm{Im}| \omega \gg 1$.
If we further impose the condition that the radiative corrections should not exceed the light neutrino masses, we find the limit on the parameter $\Mu$ in the relation \eqref{SymmetryBreaking},
\begin{align}
	\mu \lesssim  \frac{\sqrt{\sum m_i^2}}{\bar{M} U^2}
	\frac{2 \sqrt{2} \pi^2 v^2}{ \bar{M}^2 l^\prime (\bar{M}^2)}\,.
\end{align}
In the regime $\bar{M}\gg v$ where this bound turns out to be most relevant the loop function is roughly constant, and the above relation simplifies to
\begin{align}
	\mu \lesssim  \frac{\sqrt{\sum m_i^2}}{\bar{M} U^2}
	\frac{4 \sqrt{2} \pi^2 v^2}{m_H^2 +3 m_Z^2}\approx
	40 \frac{\sqrt{\sum m_i^2}}{\bar{M} U^2}\,.
\end{align}
We can combine this with the condition $\mathcal{R}_{\ell\ell}<1/3$
to find that the ratio $\mathcal{R}_{\ell \ell}$ is suppressed for
\begin{align}
	U^2 > 90 \times \frac{v}{\M} \frac{\sqrt[4]{\sum_i m_i^2}}{\sqrt{\M}} \approx 0.1 \left( \frac{\bar{M}}{\rm GeV} \right)^{-3/2}\,.
\label{LNVcondition}
\end{align}

\section{Models with more than two heavy neutrinos}
If there are more than two heavy neutrinos one can expect that the above statements are relaxed due to the larger dimensionality of the parameter space.
In the following we briefly and qualitatively discuss how the conclusions change with $n>2$, We provide a more quantitative discussion in appendix \ref{app:NHNL}.

We shall first state that the discussion below only applies to heavy neutrinos that make a significant contribution to the generation of the light neutrino masses. 
It is clear that one can always add an arbitrary number of heavy neutrinos that either have very tiny Yukawa couplings $F_{ai}\ll \sqrt{m_i M_i}/v$ or come in pseudo-Dirac pairs with 
deviations from the relation \eqref{eq:symmetry} that are too tiny to lead to a measurable contribution to the $m_i$.
In the former case those will not be visible, and in the latter case they will (almost) exclusively lead to LNC decays because they would effectively be Dirac particles.

One can generically expect the bounds to weaken in a way that scales linear with $n$ because the burden to explain the light neutrino masses is shared amongst the different pairs, but how exactly this happens is model dependent.
However, a qualitative difference arises from the fact that the mass $m_{\rm lightest}$ of the lightest SM neutrino can be non-zero for $n>2$.
If $m_{\rm lightest} > \Delta M_{\theta\theta}$, then $m_{\rm lightest}$ fixes the relevant mass scale that $\Delta M_{\rm phys}$ should be compared to.

A special case occurs if $N_i$ that are \emph{not} part of the same pseudo-Dirac pair have degenerate masses. 
Then a small perturbation of the relevant sub-matrix in $M_M$ from unity can cause large mixings among all those heavy neutrinos.
Such a scenario can lead to observable LNV at colliders for even larger mixing angles than in the $n=2$ case.
For example, consider $n=3$ with
\begin{align}
	M_M \approx \bar{M}
	\begin{pmatrix}
		1 & 0 & \muup\\
		0 & 1 & \ii \muup\\
		\muup & \ii \muup & 1
	\end{pmatrix} \ ,
	\
	F = 
	\begin{pmatrix}
		F_e & \ii F_e & 0\\
		F_\mu & \ii F_\mu & 0\\
		F_\tau & \ii F_\tau & 0
	\end{pmatrix}\,.
	\label{MNcommunism}
\end{align}
The light neutrino masses vanish at tree level, $m_\nu^{\rm tree}=0$.
The heavy neutrino mass depends on both the small parameter $\muup \ll 1$ and the mixing angle $U^2\ll1$.
The mass spectrum of the heavy neutrinos is given by
\begin{align}
M_N^\mathrm{diag} \simeq \M\mathrm{diag}(1+U^2/2, 1 + U^2/4 + \sqrt{\muup^2 + U^4/16} ,1 + U^2/4 - \sqrt{\muup^2 + U^4/16})\,.
\end{align}
For $1\gg U^2 \gg \muup$, the mass splitting between the pseudo-Dirac pair is given by $2 \muup^2/U^2$.
In the opposite scenario, where $1\gg \muup \gg U^2$, the degenearacy between the three heavy neutrinos is broken by $\muup$, and we can approximate the heavy neutrino spectrum as
$M_N^\mathrm{diag} \simeq \M\mathrm{diag}(1 + \muup, 1 ,1 - \muup)$.
Furthermore, the mixing angles of all three neutrinos are of a similar size
$|\theta_{ai}|^2\simeq v^2/\bar{M}^2|F_a|^2(1,1/2,1/2)$, hence all three heavy neutrinos could be produced at a collider.

To first order in $\muup$, the radiative corrections to the light neutrino masses vanish.
However, the mass splitting parameter $\muup$ can still not be arbitrarily large, as the $\mathcal{O}(\Delta M_N^2)$ corrections remain finite, which gives
\begin{align}
	(m_\nu^\text{1-loop})_{ab} &\approx
	-\frac{2}{(4\pi v)^2} \frac{l''(\bar{M}^2)}{2} \theta_{a i} (M_N^\dagger M_N - \bar{M}^2)^2_{ij} \theta_{bj} \bar{M}\\\notag
	&= -\frac{2}{(4\pi v)^2} \frac{l''(\bar{M}^2)}{2} \theta_a \theta_b \bar{M}^5 \times 4{\muup}^2\,.
\end{align}
The higher power of $\muup$ leads to a weaker limit on the mass splitting than in the case with two heavy neutrinos 
\begin{align}
	{\muup}^2 \lesssim
	\frac{\sqrt{\sum_i m_{\nu i}^2}}{\bar{M} U^2}
	\frac{(4 \pi v)^2}{m_H^2+3 m_Z^2}\,,
\end{align}
where we approximated the loop function for $\bar{M}\gg m_H$.
In turn, the condition that $\mathcal{R}_{\ell\ell}<1/3$ also becomes weaker
\begin{align}
	U^2>
	\begin{cases}
		30 (\sum_i m_i^2)^{1/6} v^{2/3}/\M \approx 0.4 ( \mathrm{GeV}/\M)\, \text{for}\, m_H < \M < 1\,\mathrm{TeV}\,,\\
		170 \left( \frac{v^4}{\bar{M}^4} \frac{\sqrt{\sum_i m_{i}^2}}{\bar{M}} \right)^{1/3}
		\approx 70 \left(\frac{\bar{M}}{\rm GeV} \right)^{-5/3}\, \text{for}\, \M > 1\,\mathrm{TeV}\,,
	\end{cases}
\end{align}
where the exact pre-factor depends on the value of $m_{\rm lightest}$. In this estimate we have assumed that the sum of the light neutrino masses does not exceed the cosmological bound~\cite{Aghanim:2018eyx}.
Note that the $R_{\ell\ell}=1/3$ line falls in the $U^2 > \muup$ region for masses below $1$ TeV, while for larger masses we have $U^2 < \muup$.

\section{Discussion and Conclusions}
We studied the perspectives to see LNV signatures in collider searches for heavy neutrinos $N_i$
that interact with the SM exclusively via their mixings $\theta_{ai}$ with the SM neutrinos, where $a$ denotes a SM generation.
Searches for heavy neutrinos with Majorana masses $M_i$ below the TeV scale are highly motivated because they
can simultaneously explain the light neutrino oscillation data and the baryon asymmetry of the universe. 
Mixing angles $\theta_{ai}$ that are large enough to yield observable production cross sections at the LHC can be made consistent with the smallness of the light neutrino masses if the heavy neutrino interactions respect the approximate conservation of a generalised lepton number $\bar{L}$.
The underlying symmetry protects the light neutrino masses from large corrections even though the mixing angles $\theta_{ai} \sim F_a v/M_i$ are much larger than the naive seesaw relation \eqref{eq:seesaw} would suggest, which is a necessary requirement for the $N_i$ to be produced in sizeable numbers at the LHC.

On the other hand the $\bar{L}$ conservation can potentially suppress the branching ratios of LNV processes at colliders.
The symmetry dictates that the heavy neutrinos must be organised in pairs with quasi-degenerate masses.
Whether or not the suppression of LNV decays is efficient crucially relies on the ratio between the mass splittings that break the mass degeneracy and the decay widths of the heavy neutrinos in each pair. 
If the physical mass splitting is smaller than the decay width, then the branching ratio $\mathcal{R}_{\ell\ell}$ between LNV and LNC processes is parametrically suppressed due to destructive interference between the contributions from the two heavy neutrinos.
If the mass splitting is  larger than the decay width, then the heavy neutrinos undergo many oscillations in the detector before they decay.
In this case LNV processes are unsuppressed due to the effective loss of  quantum coherence.

By comparing the light and heavy neutrino mass matrices \eqref{eq:seesaw} and \eqref{MNdef} we argue that, leaving aside (tuned) accidental cancellations, the physical heavy neutrino mass splitting should at least be of the same order as the light neutrino mass splittings. 
This allows us to define a line in the heavy neutrino mass-mixing plane below which the violation of $\bar{L}$-conservation that is required to generate the light neutrino masses generically induces LNV heavy neutrino decays with a branching ratio $\mathcal{R}_{\ell\ell}$ of order unity.

At the same time one can impose an upper bound on the heavy neutrino mass splitting from the requirement that the light neutrino masses are stable under radiative corrections. 
Since this upper bound depends on the heavy neutrino mixings, this defines a second line in the heavy neutrino mass-mixing plane above which $\mathcal{R}_{\ell\ell}$ is generically suppressed. 
Deviations from the generic behaviour in either region can only be realised for fine tuned parameter choices that violate the criterion of technical naturalness.
In between there is a third region of masses and mixings where technically natural parameter choices that do or do not give a $\mathcal{R}_{\ell\ell}$ or order unity can coexist. 

The precise locations of these regions depend on the number of right handed neutrinos $n$ that contribute to the generation of light neutrino masses. 
We give parametric estimates of the boundaries in eqns.~\eqref{superEWcond}, \eqref{subEWcond} and \eqref{largetheta2radiative}, respectively.
In figure \ref{fig:LNVboundary} we display the three regions for the minimal model with $n=2$. 
Roughly speaking, for heavy neutrino masses below the electroweak scale the observation of LNV processes is the rule rather than the exception in the parameter region that is allowed by current experimental limits.
For masses above the electroweak scale, on the other hand, $\mathcal{R}_{\ell\ell}$ is suppressed. Using the simplified scalings $\mathcal{R}_{\ell\ell}\propto U^{-4}$ and $\sigma_N \propto U^2$ one can estimate that the suppression in the number of LNV events is quadratic in the mixing angle.

In models with more than two heavy neutrinos the constraints derived here relax due to the higher dimensionality of the parameter space, in particular if there are accidental mass degeneracies in addition to those dictated by the $\bar{L}$-symmetry.
Finally we should emphasise again that our analysis only applies to models in which the heavy neutrinos primarily interact via their Yukawa couplings and generate the light neutrino masses in this manner. If the mixing between left handed and right handed neutrinos is not the sole origin of neutrino mass or if the right handed neutrinos have new interactions that are relevant at energies below the TeV scale then there are additional ways to make LNV observable.

We can therefore conclude that searches for LNV processes mediated by heavy neutrino exchange are well-motivated across the entire experimentally accessible mass range. 
Studying the ratio between LNV and LNC decay rates as well as the angular distribution of the decay products can give interesting insight into the underlying mechanism of neutrino mass generation.

\section*{Acknowledgements}
We thank Björn Garbrecht, Richard Ruiz and Inar Timiryasov for helpful discussions.
M.D. is grateful to the Max Planck Institute for Physics (Werner Heisenberg Institut) in Munich for their hospitality during the work on this project.
J.K. acknowledges the support of the the ERC-AdG-2015 grant 694896. 

\appendix
\section{Stability of light neutrino masses under radiative corrections}
\label{app:NHNL}
In ref.~\cite{Moffat:2017feq} it was shown that massless neutrinos are stable under radiative corrections only if the Yukawa couplings take the form from~\eqref{eq:symmetry} or vanish.
To obtain small but finite neutrino masses this pattern has to be mildly broken by terms that are parametrically small.
In this appendix we quantify the allowed size of these symmetry breaking parameters, and hence the amount of LNV that can be observed at colliders.

Let us first review the case with two heavy neutrinos. In the seesaw limit the light neutrino mass matrix takes the form
\begin{align}
	(m^{\rm tree}_\nu)_{ab} = - \sum_i \theta_{a i} \theta_{b i} M_i\,,
	\label{eq:treelvl2HNL}
\end{align}
For fixed values of the mixing angles $\theta_{a 2}$ as well as the masses $M_1$ and $M_2$, one can satisfy the above equation by solving it for $\theta_{a1}$.
On the other hand, one still has to make sure that the radiative corrections are not bigger than the light neutrino masses.
Since $\theta_{a1}$ is already fixed by solving~\eqref{eq:treelvl2HNL}, the size of the radiative corrections is completely determined by the mixing $\theta_{a2}$, and the masses $M_1$ and $M_2$,
\begin{align}
	(m_\nu^{1-\mathrm{loop}})_{ab} &= \frac{2}{(4 \pi v)^2}
	\left\{ (m_\nu^\mathrm{tree})_{ab} l(M_1^2)
		- \sum_{i>1} \theta_{ai} \theta_{bi} M_i [l(M_i^2) - l(M_1^2)]
	\right\}\,,
	\label{eq:generalN}\\
	&= \frac{2}{(4 \pi v)^2}
	\left\{ (m_\nu^\mathrm{tree})_{ab} l(M_1^2)
		- \theta_{a2} \theta_{b2} M_2 [l(M_2^2) - l(M_1^2)]
	\right\}\,,
	\label{eq:radiativew2HNL}
\end{align}
which gives the constraint~\eqref{ns2radiativecorr}.

If one attempts the same exercise with $n>2$, one finds that there are enough parameters to make the radiative corrections vanish, e.g. by choosing a particular value of $\theta_{a3}$.
However, this choice will explicitly depend on the loop function $l$, and the mass scales appearing within.
To relate this dependence to a small parameter we perform an expansion of $l(M_i)$ around $l(M_1)$
\begin{align}
	(m_\nu^{1-\mathrm{loop}})_{ab} &=\frac{2}{(4 \pi v)^2}
	\left\{ (m_\nu^\mathrm{tree})_{ab} l(M_1^2)
		- \sum_{i>1,n} \theta_{ai} \theta_{bi} \frac{1}{n!} \frac{d^n l(M_1^2)}{d (M_1^2)^n} M_i [M_i^2 - M_1^2]^n
	\right\}\,.
\end{align}
As we wish to avoid explicit dependence on $l$, we treat each order in this series as an independent equation.
This infinite tower of equations can never be solved unless the mass differences exactly vanish.
However, for $n$ heavy neutrinos, we can solve these equations up to order $(M_i-M_1)^{n-1}$, which leads to a quantitatively weaker limit on the mass splitting:
\begin{align}
	\sqrt{\sum_i m_{i}^2} &\gtrsim
	\frac{2}{(4\pi v)^2}\frac{1}{(n-1)!} \frac{d^{n-1} l(M_1^2)}{d (M_1^2)^{N-1}} U_N^2 M_1^{2n +1} \mu^{n-1}\,,\\\notag
	&\gtrsim
	\frac{m_H^2+3 m_Z^2}{(4 \pi v)^2} M_1 U_N^2 \mu^{n-1}\,,
\end{align}
where we assumed that all the mass splittings are of a similar size $\sim \mu M_1$.

\bibliographystyle{JHEP}
\bibliography{RHNrefs}{}

\end{document}